\documentclass[final]{iopart}
\usepackage[utf8]{inputenc}
\usepackage[T1]{fontenc}
\usepackage{amsmath}
\usepackage{amssymb}
\usepackage{esvect}
\usepackage{footnote}
\usepackage{graphicx}

\usepackage[justification=justified]{subfig}
\usepackage{color}

\begin{document}

\title{Boolean networks with robust and reliable trajectories}
\author{Christoph Schmal, Tiago P. Peixoto and Barbara Drossel}
\address{Institut für Festkörperphysik, TU Darmstadt, Hochschulstrasse 6,
  64289 Darmstadt, Germany}
\eads{\mailto{schmal@physik.uni-bielefeld.de}, \mailto{tiago@fkp.tu-darmstadt.de},
  \mailto{drossel@fkp.tu-darmstadt.de}}

\begin{abstract}
  We construct and investigate Boolean networks that follow a given reliable
  trajectory in state space, which is insensitive to fluctuations in the
  updating schedule, and which is also robust against noise. Robustness is
  quantified as the probability that the dynamics return to the reliable
  trajectory after a perturbation of the state of a single node. In order to
  achieve high robustness, we navigate through the space of possible update
  functions by using an evolutionary algorithm. We constrain the networks to
  having the minimum number of connections required to obtain the reliable
  trajectory. Surprisingly, we find that robustness always reaches values close
  to 100 percent during the evolutionary optimization process. The set of update
  functions can be evolved such that it differs only slightly from that of
  networks that were not optimized with respect to robustness. The state space
  of the optimized networks is dominated by the basin of attraction of the
  reliable trajectory.
\end{abstract}

\maketitle
\section{Introduction}

Boolean networks (BNs) are used as simplified models of gene regulation,
where the expression levels of genes are described by Boolean variables,
and their mutual regulation by Boolean functions. This simplification
permits in particular the analysis of larger networks, the full dynamics
of which would include many nonlinear equations and many
parameters~\cite{bornholdt_systems_2005}.

The simplest class of BNs are Random Boolean Networks
(RBNs)~\cite{kauffman_metabolic_1969, drossel_random_2008}, i.e., BNs
with connections and update functions assigned at random to each
node. These networks undergo a phase transition from a frozen phase to a
"chaotic" phase at a critical value $K=2$ of the number of inputs per
node. It has been argued ~\cite{kauffman_metabolic_1969} that real
networks may share properties with RBNs that lie at the boundary between
two phases, since these ``critical'' networks are capable of responding
to perturbations, but without an exponentially fast divergence of
trajectories in state space.

However, critical RBNs are not robust against
noise~\cite{peixoto_noise_2009}, due to their large number of dynamical
attractors. In contrast, BNs that are modeled on the basis of real
biological data, such as the yeast cell cycle regulation
network~\cite{li_yeast_2004}, go faithfully through the correct sequence
of states even in the presence of noise. This is due to the structure of
the state space: most states of the network lead after a few update
steps to the dynamical attractor that corresponds to the cell cycle.

In this paper, we will construct and investigate BNs that are robust against two
types of noise. The first type of noise is applied to the \emph{update
  schedule}, and it delays or advances the update time of a given
node~\cite{greil_dynamics_2005,klemm_stable_2005, aracena_robustness_2009,
  peixoto_boolean_2009}.  The second type of noise is applied to the
\emph{update rule}, and it flips the state of a node to the opposite of the
value imposed by the update function~\cite{miranda_noise_1989,
  aleksiejuk_ferromagnetic_2002, huepe_dynamical_2002, qu_numerical_2002,
  indekeu_special_2004, ribeiro_noisy_2007, peixoto_noise_2009,
  serra_dynamics_2010}. Both types of noise are present in real systems, since
genes lack a global update clock and are therefore not updated at fixed time
intervals, and since expression levels are subject to stochastic
fluctuation~\cite{mcadams_stochastic_1997}. However, these two types of noise
are quite different, and require different strategies to attain robustness: With
respect to the first type of noise, it is possible for the dynamics of BNs to be
entirely reliable, simply by requiring that consecutive states of an attractor
differ by the state of at most one node~\cite{peixoto_boolean_2009}. In order to
make BNs robust against the second type of noise, it is necessary to introduce
redundancy~\cite{peixoto_redundancy_2010}, or to build networks with a state
space dominated by the basin of attraction of one
attractor~\cite{szejka_evolution_2007}. These methods lead to a good level of
robustness, but can never entirely remove the effects of noise.

In order to obtain networks that are robust against both types of noise, we will
first construct minimal networks that have a \emph{reliable} dynamical
trajectory, which is insensitive to the sequence in which the nodes are
updated~\cite{peixoto_boolean_2009}. Then, by using an evolutionary algorithm,
we will optimize the set of update functions of all nodes such that the dynamics
return to this attractor with a large probability when the state of a node is
perturbed.  We investigate the extent of robustness attainable for these
networks, and characterize the distribution of their update functions and their
state space properties.

This paper is structured as follows: In section~\ref{sec:model}, we
provide a definition of the model, and a description of the minimal
reliable BNs defined in~\cite{peixoto_boolean_2009} as well as of the
evolutionary algorithm used for the optimization process. In
section~\ref{sec:results}, we analyze the robustness, the set of
update functions, and the state space of the networks obtained by the
optimization process.  Section~\ref{sec:conclusion} summarizes and
discusses our main findings.

\section{Construction of reliable and robust BNs} \label{sec:model}

Our goal is to obtain and investigate BNs that are robust with respect to the
update schedule and with respect to perturbation of the state of a node. To this
purpose, we first construct reliable networks (i.e., networks that have an
attractor that is robust with respect to the update sequence), which we will
then optimize with respect to robustness against perturbations.

\subsection{Reliable BNs}

A Boolean network is specified by its
topology and dynamical update rules. The topology is specified by the number
$N$ of nodes, and by the connections between these nodes. Each node obtains an index  $i\in \{0,
1,..., N-1\}$ and can be either in the state $\sigma_i=0$ or $\sigma_i=1$. Its
time evolution is given by the iterative map
\begin{align}
 \sigma_i(t+1) = f_i(\vv{\sigma}_{j(i)}(t))u_i(t)+\sigma_i(t)[1-u_i(t)]
\end{align}
where $f_i:\{0,1\}^{k_i}\mapsto \{0,1\}$ is the update function of node $i$,
which depends exclusively on the states of its $k_i$ input nodes
$\vv{\sigma}_{j(i)}$. $\vv{u}(t)$ represents the update schedule, and has the
components $u_i(t)=1$ if node $i$ is updated at time $t$, and $u_i(t)=0$ if it
is not updated at time $t$. 

We construct networks with entirely reliable trajectories in the same
way as in~\cite{peixoto_boolean_2009}. Reliable trajectories have the
property that two consecutive states (under any update schedule)
$\vv{\sigma}(t)$ and $\vv{\sigma}(t+1)$ can differ by the
value of at most one node, i.e. the Hamming distance between these
states is one.
\begin{figure}[htbp]
\centering
 \includegraphics[width=0.3\linewidth]{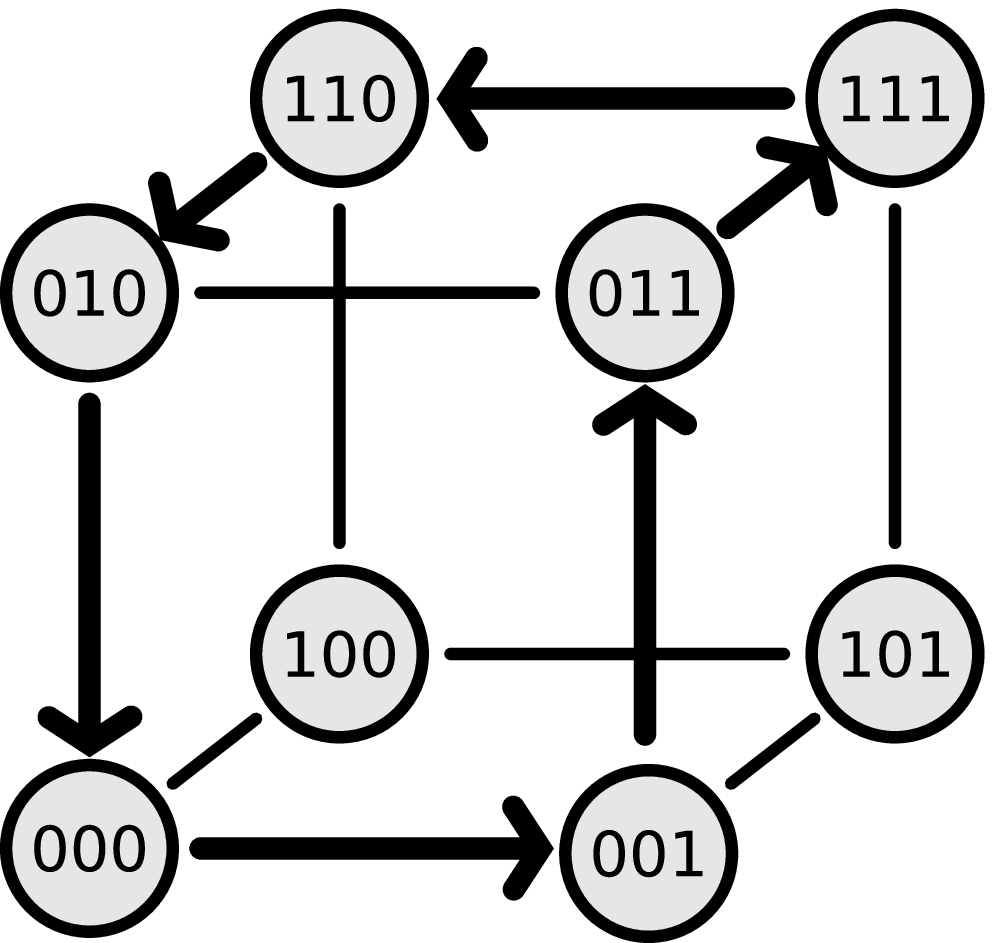}
 \caption{Example of a reliable trajectory of length $L=6$ on a system of size $N=3$.}
 \label{fig:hammingcube}
\end{figure}
Entirely reliable attractors can therefore be represented as closed
walks over the $N$-dimensional Hamming hypercube, as shown in
Fig.~\ref{fig:hammingcube}. The length of the attractor can be written
as $L=\sum_il_i$, where $l_i$ denotes the number of times node $i$
changes its state during the full period. Given a reliable trajectory
of length $L$ it is possible to construct a \emph{minimal} network that realizes
it, by finding for each node a minimal set of inputs and a corresponding
Boolean function which is compatible with the trajectory (see
\cite{peixoto_boolean_2009} for details). Since there are possibly
many such networks, we sample randomly from the ensemble of all
possibilities. From all possible functions that realize the same
trajectory, given a specific choice of inputs, we choose those which
are more \emph{homogeneous}, i.e., that have the smallest number of
outputs that deviate from the majority bit in their truth table.

\begin{figure}[htbp]
  \centering
 \includegraphics[width=0.65\linewidth]{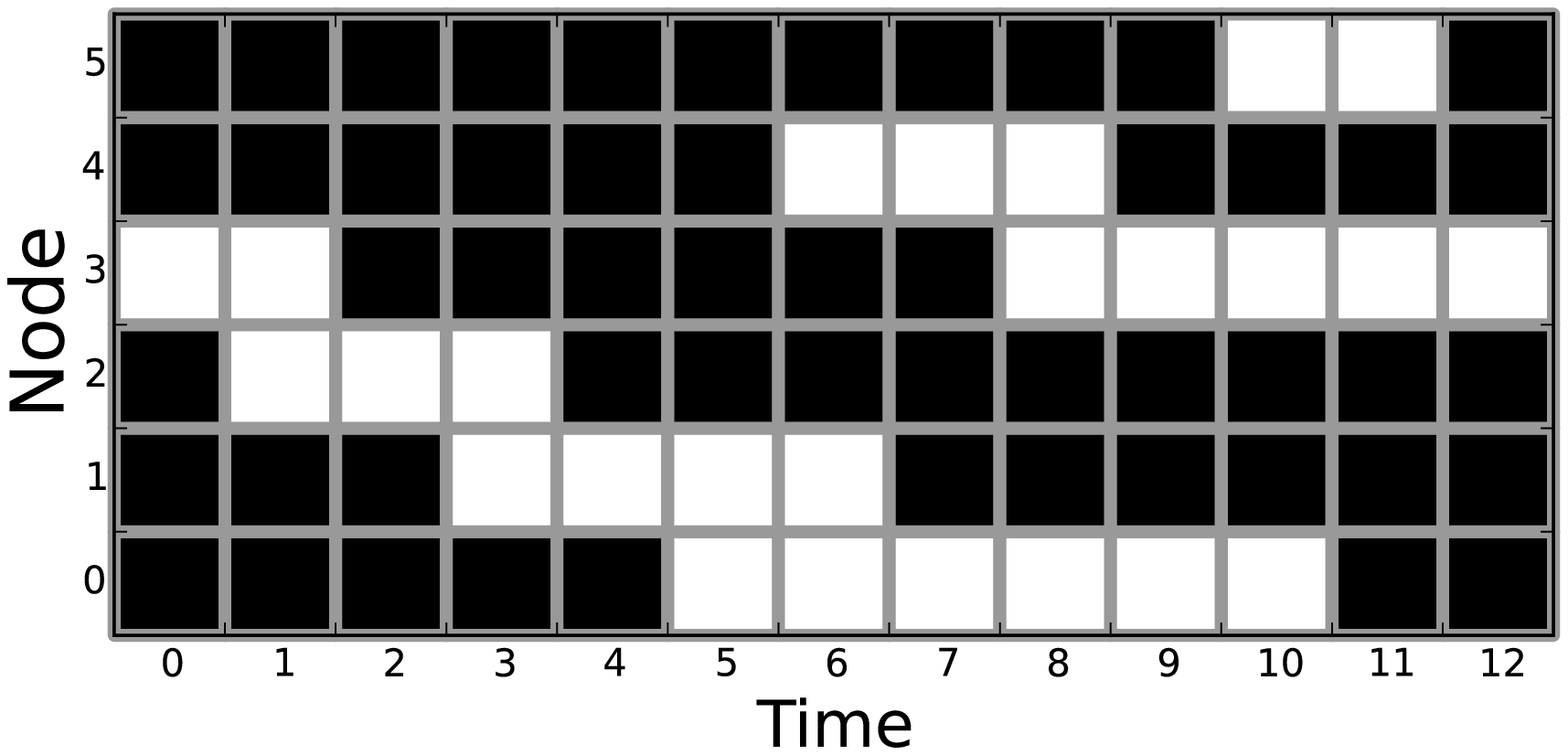}
 \includegraphics[width=0.33\linewidth]{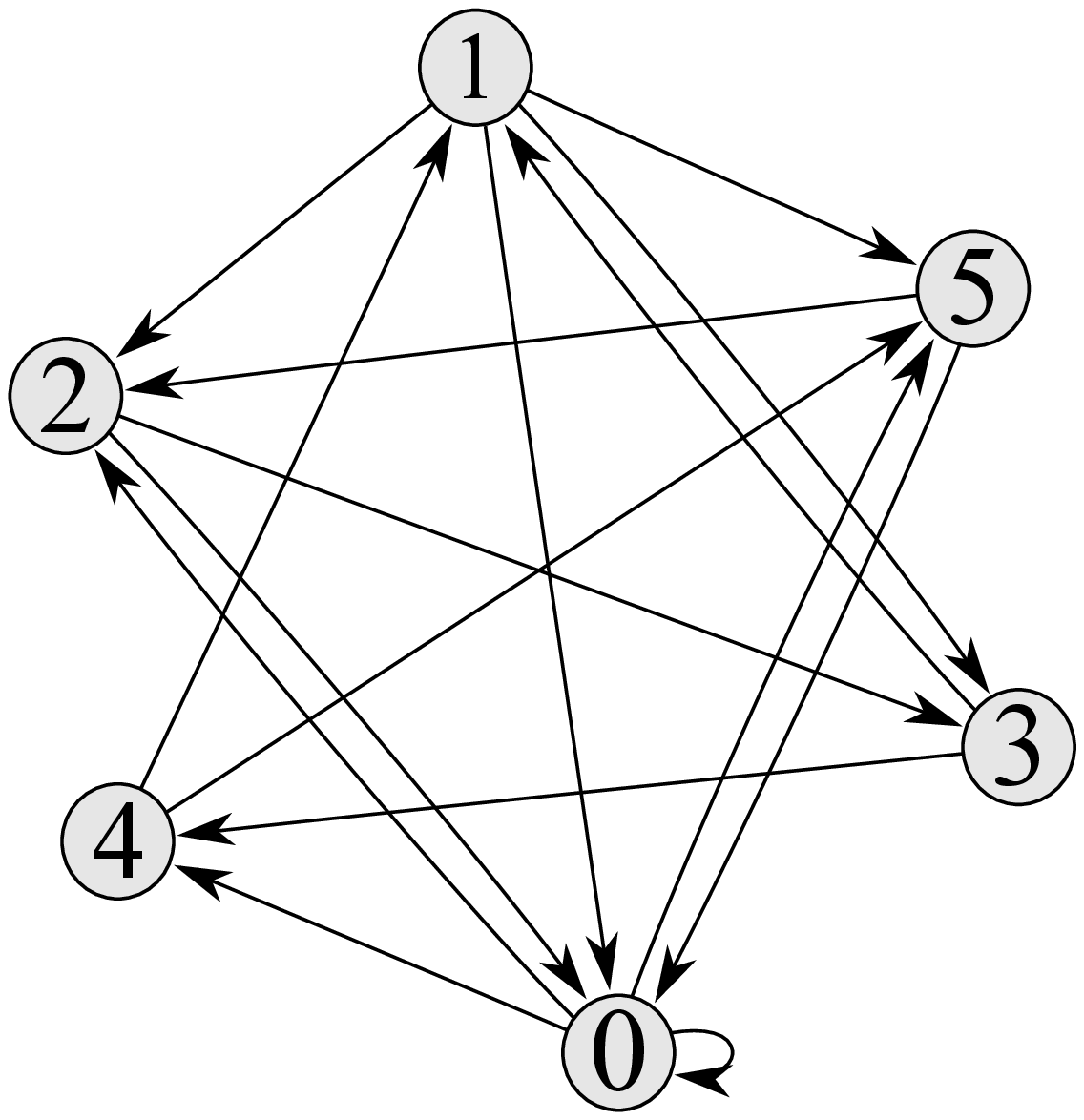}
 \caption{Example of a random reliable trajectory and one corresponding minimal
   network.}
 \label{fig:bsp_nw}
\end{figure}

We generate the reliable trajectories at random, given the average number of
flips per node $l$.  The number of flips of node $i$ is $l_i = 2 + 2\ell_i$,
where $\ell_i$ is a random variable sampled from a Poisson distribution with
average $l/2-1$. The average total length of the trajectory is given simply by
$Nl$. Fig.~\ref{fig:bsp_nw} shows an example of a random trajectory and one of
its minimal networks.

\subsection{Optimizing the networks for dynamical robustness}\label{sec:robustness}

We define robustness
as the probability that the dynamics return to the reliable trajectory after a
perturbation of a single node. Such a perturbation moves the system to one of
the $N$ neighboring states on the Hamming hypercube representing the state space.
More precisely, considering the set
\begin{align}
\mathcal{H}_1 (\vv{\sigma}_a) = \{ \vv{\sigma} \in \{0,1\}^N : H(\vv{\sigma}, \vv{\sigma}_a) = 1  \}
\end{align}
of all states with Hamming-distance 1 from a given state $\vv{\sigma}_a$ of
our reliable attractor, we define the local fitness $f_a$ of this state as the
fraction of these $N$ neighbors that return to the reliable attractor. The total
fitness of the network is given  by the average $f = \sum_{a=1}^L f_a / L$. In order to avoid
stochasticity in the measurement of $f$, we always use a parallel update
schedule, where all nodes are updated at the same time.

The fact that two successive states on the reliable trajectory differ only by
the value of one node means that there is a lower bound on the fitness value
 of $f_{\text{min}} = 2/N$, since two of the $N$ possible perturbations
 generate a state that is on the reliable attractor.

Given this definition of the fitness of the network, we apply an
evolutionary algorithm in order to maximize it, modifying the update
functions but retaining the network topology and the reliable
trajectory.  When exploring the search space $\mathcal{S}$ of possible
update functions, we can only change the truth table entries of the
output values that do not interfere with the given reliable
trajectory.  Let us assume that node $i$ has $k_i$ input nodes. If its
function has $\kappa_i$ truth table entries that are fixed by the
reliable trajectory and $\varkappa_i$ entries that are not, then there
are $2^{\varkappa_i}$ different possible output combinations for these
entries. For $N$ nodes, we have $|\mathcal{S}|=\Pi_{i=1}^N
2^{\varkappa_i}=2^{\sum_{i=1}^N\varkappa_i}$ for the size of the
search space. The typical number of entries not fixed by the reliable
trajectory scales as $\varkappa \sim 2^{\langle k \rangle} - \langle
\kappa\rangle\sim 2^l -l$. Hence the size of the search space scales as
$|\mathcal{S}|\sim 2^{N(2^l-l)}$ and therefore grows exponentially with
$N$ and superexponentially with $l$. Finding a global optimum by
searching through all update functions is possible
only for very small networks. Instead, we use an evolutionary algorithm, specified as follows:
\begin{enumerate}
 \item A node $i\in\{1,2,...,N\}$ is chosen at random.
 \item An output in the truth table of this node is chosen at random. If
   it does not belong to a configuration of the input nodes that occurs during
   the course of the reliable trajectory, we change its value.
 \item When this \textit{mutation} increases the fitness (\textit{positive
     mutation}) or has no effect (\textit{neutral mutation}) we accept
   the modification, otherwise (\textit{negative mutations}) we reject it. 
 \item The \emph{adaptive walk} obtained by iterating steps 1~to~3 stops when
   the maximum possible fitness value (evaluated below) is reached, or after a
   certain number of attempted mutations, which was set to $5\times10^3$ for
   $N=10$, to $10\times10^3$ for $N=20$, and to $30\times10^3$ for simulations
   that use the approximate fitness $f^{\star}$ (see below).
 \end{enumerate}

 \section{Results} \label{sec:results}
\subsection{Robustness of reliable networks before evolution}
\begin{figure}[htbp]
\centering
 \includegraphics[width=0.8\linewidth]{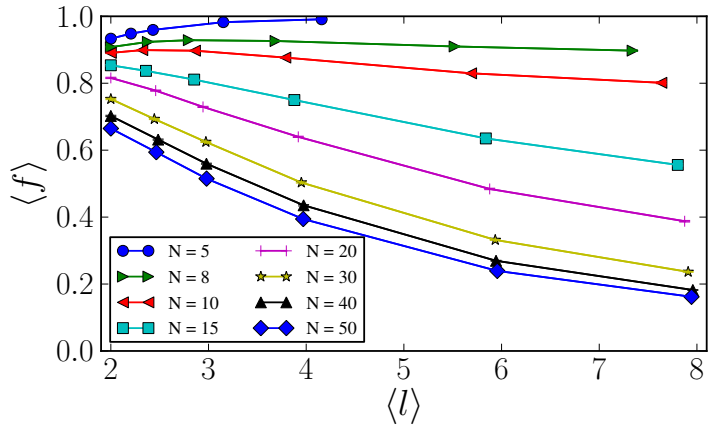}
 \caption{Average fitness $\langle f\rangle$, for several parameter
   combinations.}
 \label{fig:initial_fitness}
\end{figure}

Fig.~\ref{fig:initial_fitness} shows the initial fitness $f$ of minimal reliable networks for several
combinations of $N$ and $l$, averaged for $6\times10^3$ (for $N<40$) or $2\times10^3$ (for $N> 40$) 
independent network realizations.

A large proportion of networks with small $N$ and $l$ have $f=1$. As
was observed in~\cite{peixoto_boolean_2009}, for these networks the
reliable trajectory often has a basin of attraction which dominates the
entire state space, which explains why $f$ is close to 1. When $N$ and
$l$ increase, this changes, and the basin of the reliable trajectory
no longer dominates the state space, resulting in smaller values of
$\left<f\right>$. The only trivial exception is when $l$ is so large
that the reliable trajectory occupies a large portion of the state
space (i.e. $Nl\sim 2^N$). This explains the positive slope of the
curve for $N=5$.  In the more interesting case $N\gg 1$ and $Nl\ll
2^N$, the fitness is far from the maximum value, and the
optimization procedure can considerably increase the fitness.

\subsection{Fitness of the optimized networks}
\subsubsection{Upper bound on the fitness}
\begin{figure}[htbp]
\centering
 \subfloat[before evolution]{\includegraphics[width=0.49\linewidth]{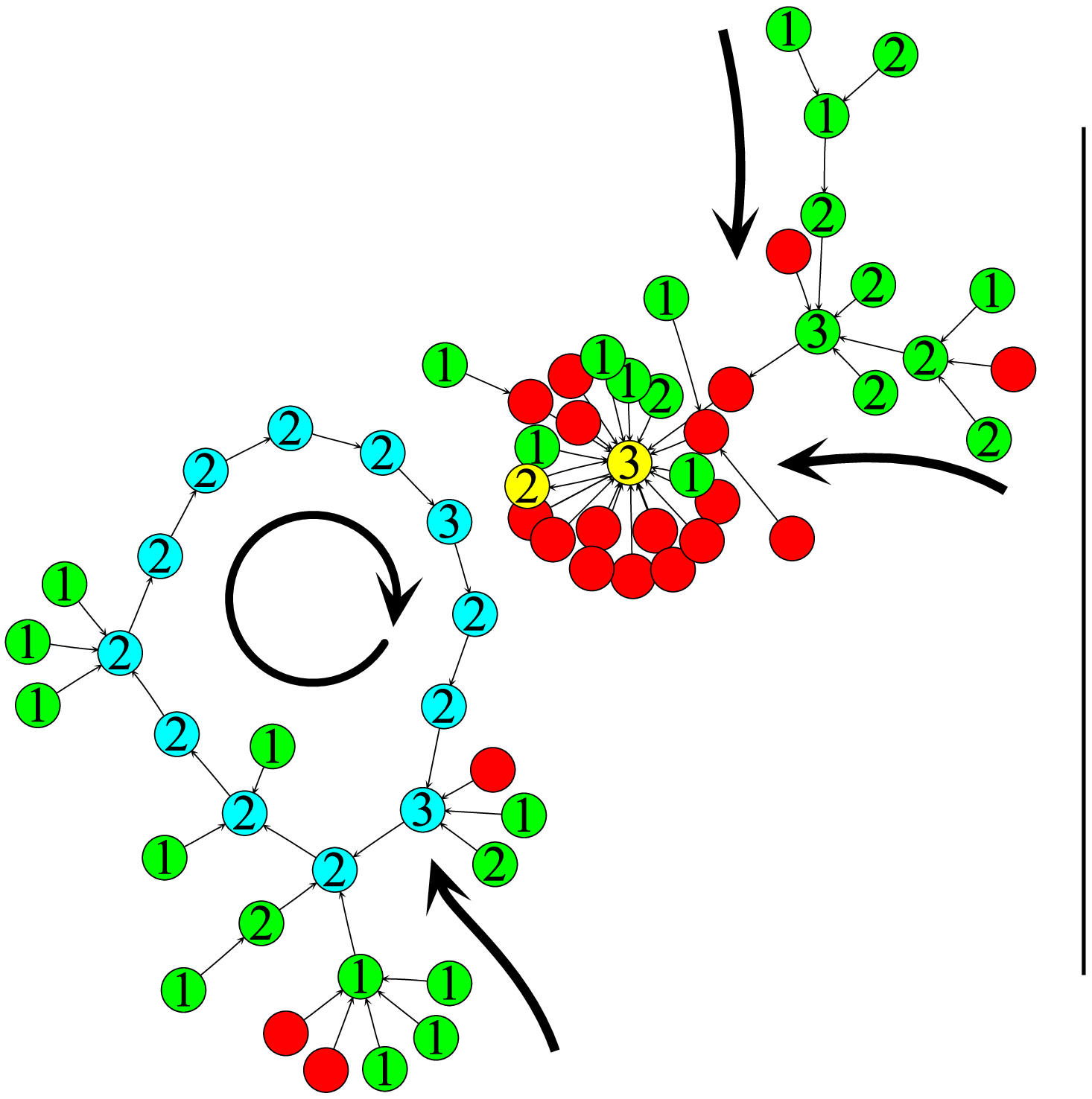}}
 \subfloat[after evolution]{\includegraphics[width=0.49\linewidth]{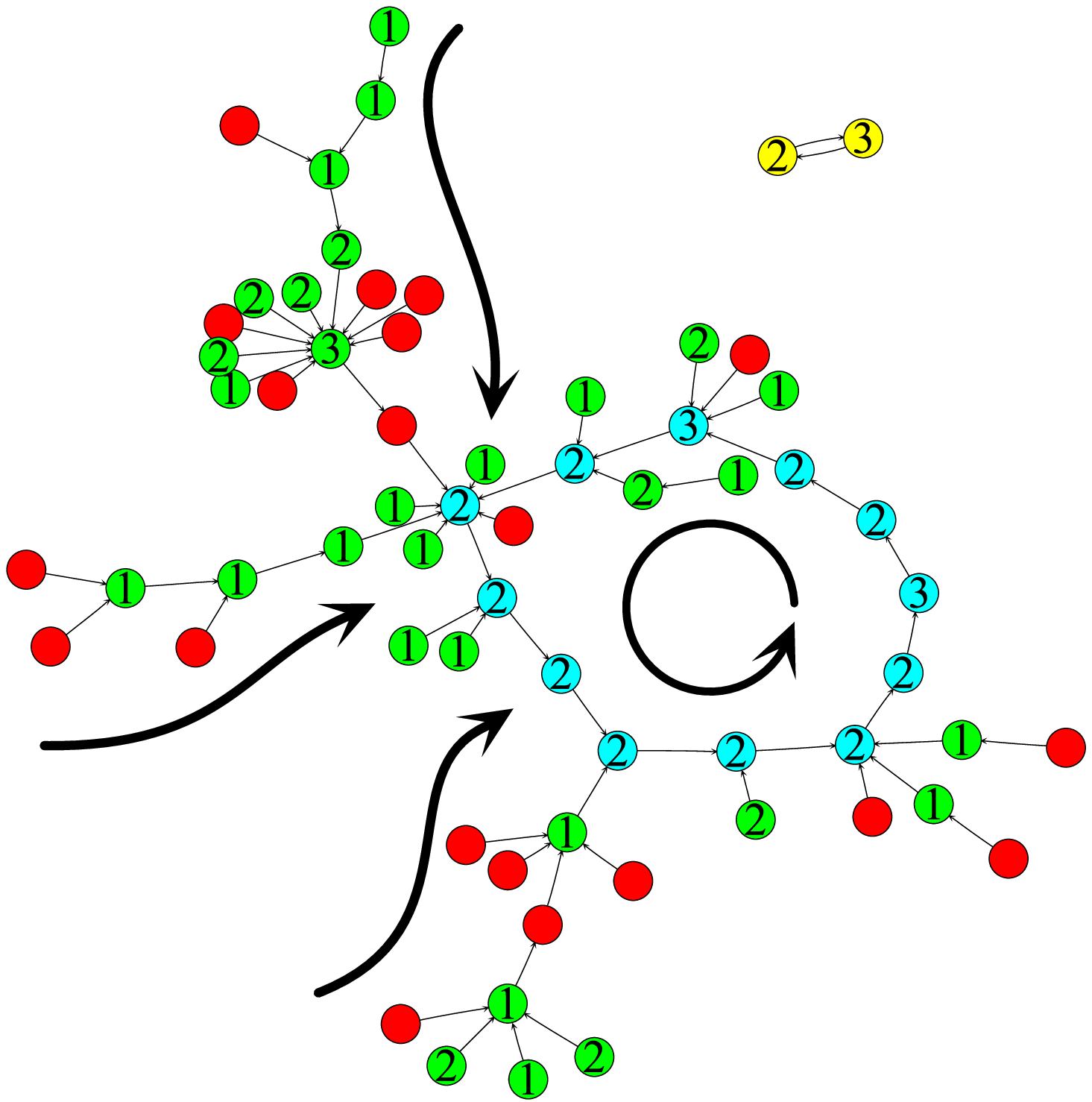}}
 \caption{State space of the example network in Fig~\ref{fig:bsp_nw} before and
   after evolution. The states are color-coded as follows. Blue: reliable
   attractor. Green: states to which the network is brought by a
   perturbation. Yellow: the attractor that cannot be modified by the
   optimization procedure and is reached by a perturbation. Red: remaining
   states.}
 \label{fig:sp_bsp_nw}
\end{figure}

\begin{figure}[htbp]
\centering
 \includegraphics[width=0.9\linewidth]{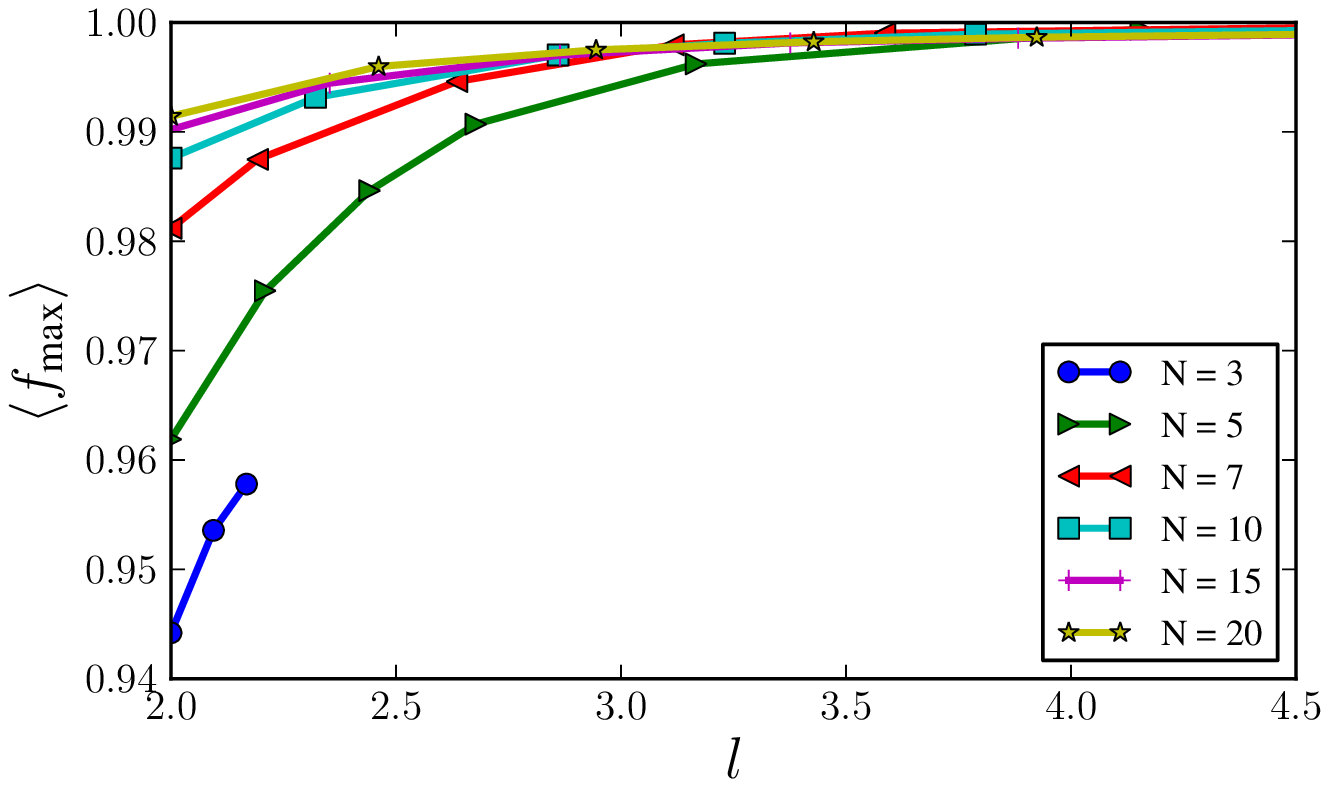}
 \includegraphics[width=0.9\linewidth]{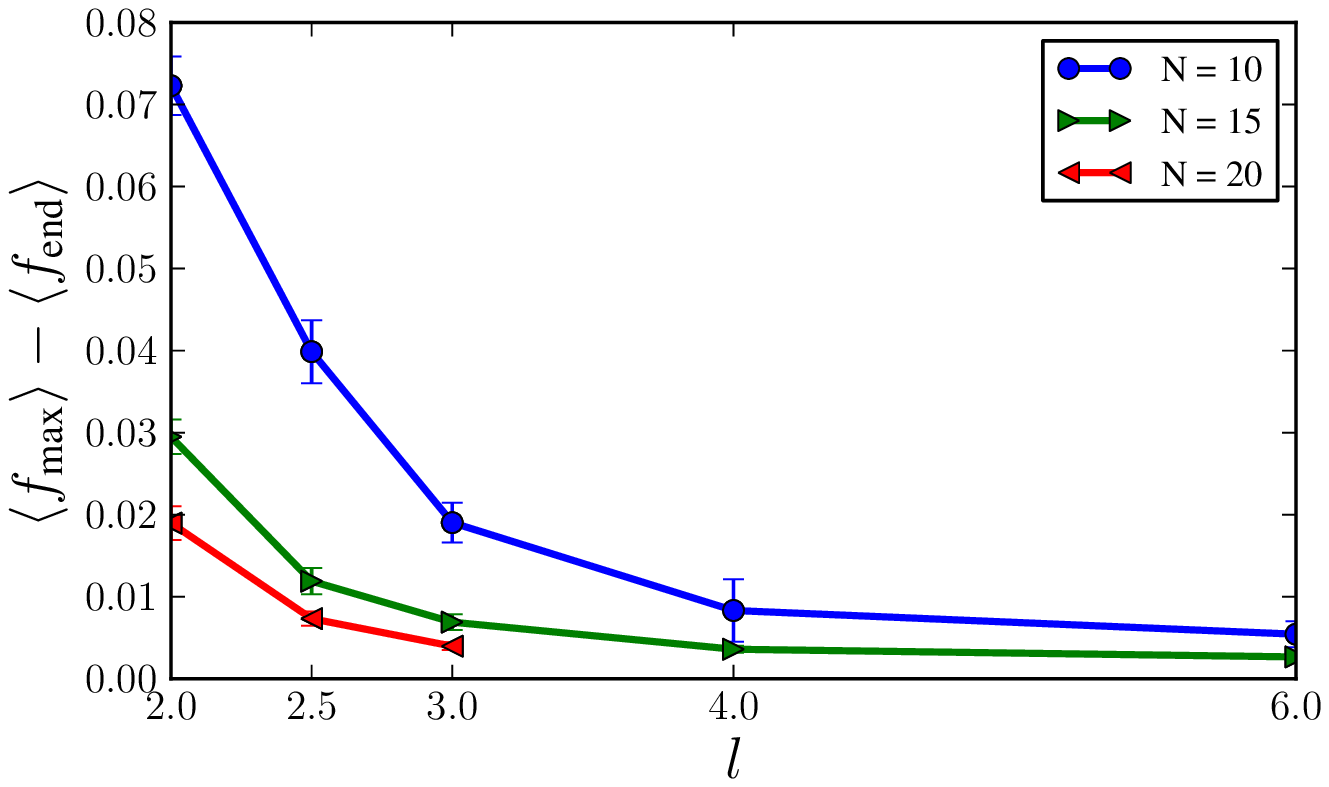}
 \includegraphics[width=0.9\linewidth]{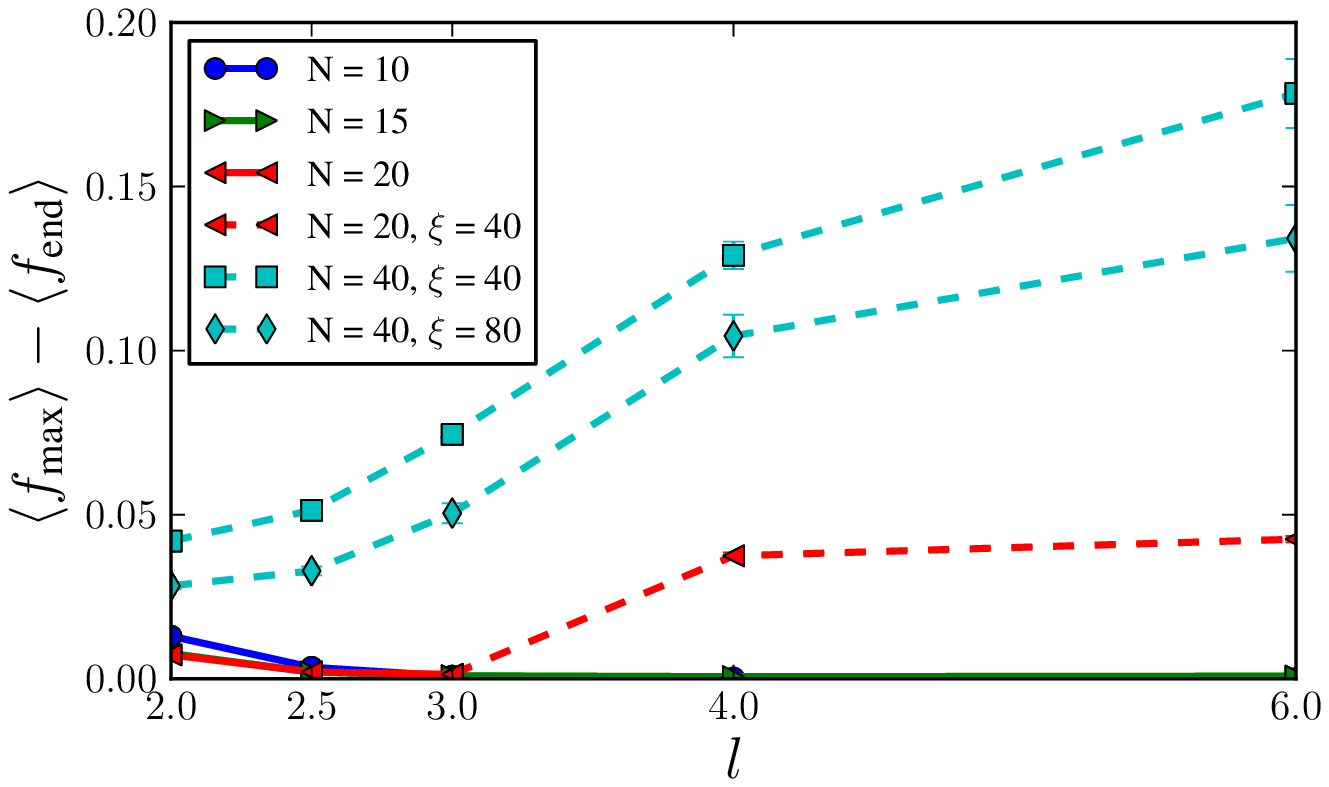}
\caption{Upper limit on fitness, averaged over networks of different sizes $N$
  and trajectory lengths $l$ (top). Average deviation from $f_{\text{max}}$
  after the evolutionary process for those networks that did not reach
  $f_{\text{max}}$ (middle).  Average deviation from $f_{\text{max}}$ after the
  evolutionary process for all networks; the dashed curves being obtained by
  evolving the functions based on the approximate fitness $f^{\star}$ (bottom).}
 \label{fig:fitness}
\end{figure}

In contrast to our initial expectation, even a full search of the
space of update functions does not always lead to a fitness value of
1.  The reason for this is that the search space is constrained by the
reliable trajectory, which cannot change during the evolutionary
algorithm. This means that the truth table entries that cannot be
modified by the evolutionary algorithm (since they are necessary to
regulate the given trajectory) may also regulate other portions of the
state space. This portion, therefore, cannot be modified by the
optimization. If some of these states are reached after a
perturbation, and they do not inherently lead back to the reliable
trajectory, then the value of $f=1$ can never be reached. If $\phi$ is
the number of perturbations which lead to one of these ``locked''
states, the maximum fitness will then be
$f_{\text{max}}=1-\frac{\phi}{NL}$.  Fig.~\ref{fig:sp_bsp_nw} shows
the state space of such a network with a maximal fitness smaller than
$f=1$. For five possible perturbations of the reliable attractor, this
network will unavoidably be trapped in a spurious attractor of size
two, and thus $f_{\text{max}} = 1-\frac{5}{NL}=67/72 \approx 0.93$.
We evaluated $f_{\text{max}}$ for ensembles of networks with different
$l$ and $N$, and observed that $f_{\text{max}}$ converges fast to 1 as
$l$ and $N$ increase, as can be seen in Fig.~\ref{fig:fitness}.  This
is easy to explain, since the typical truth table size scales as $\sim
2^l$, and the amount of unevolvable truth table entries per node
scales only as $\sim l$.  Thus the probability that,
after a perturbation, the state of a node will be regulated via such
an unvolvable truth table entry is $\sim l/2^l$.  The probability that
this will happen simultaneously for all nodes is given by $\sim
\prod_{i=1}^N p_i=\left(\frac{l}{2^l} \right)^N$, which tends to
zero for either $l\gg 1$ or $N\gg 1$.

\subsubsection{Approximate fitness}

The computer time required for the fitness evaluation at each evolutionary step
depends of the number~$L\approx Nl$ of states in the reliable trajectory, and on
the number $N$ of possible perturbations per state, which leads to a complexity
of $\mathcal{O}(N^2l)$. Thus, the optimization process becomes computationally
too expensive for larger $N$ and $l$ as we have to determine the fitness after
each mutation. In order to reduce computer time for larger $Nl$, we used an
approximate fitness function $f^{\star}$, which uses only a random subset of
$\xi$ different perturbations, which remains fixed during the
optimization. Thus, if $k$ of these $\xi$ perturbed states return to the
reliable trajectory the approximate fitness is $f^{\star}:=k/\xi$. Such an
approximation introduces a probability of accepting a negative mutation or
rejecting a positive or neutral one. In order to minimize this effect, we
re-sample the $\xi$ perturbations after the maximum $f^{\star}_{\text{max}}$ has
been reached (which can be computed analogously to $f_{\text{max}}$ above).

\begin{figure}[htbp]
  \centering
  \subfloat[$\xi=20$]{\includegraphics[width=0.49\linewidth]{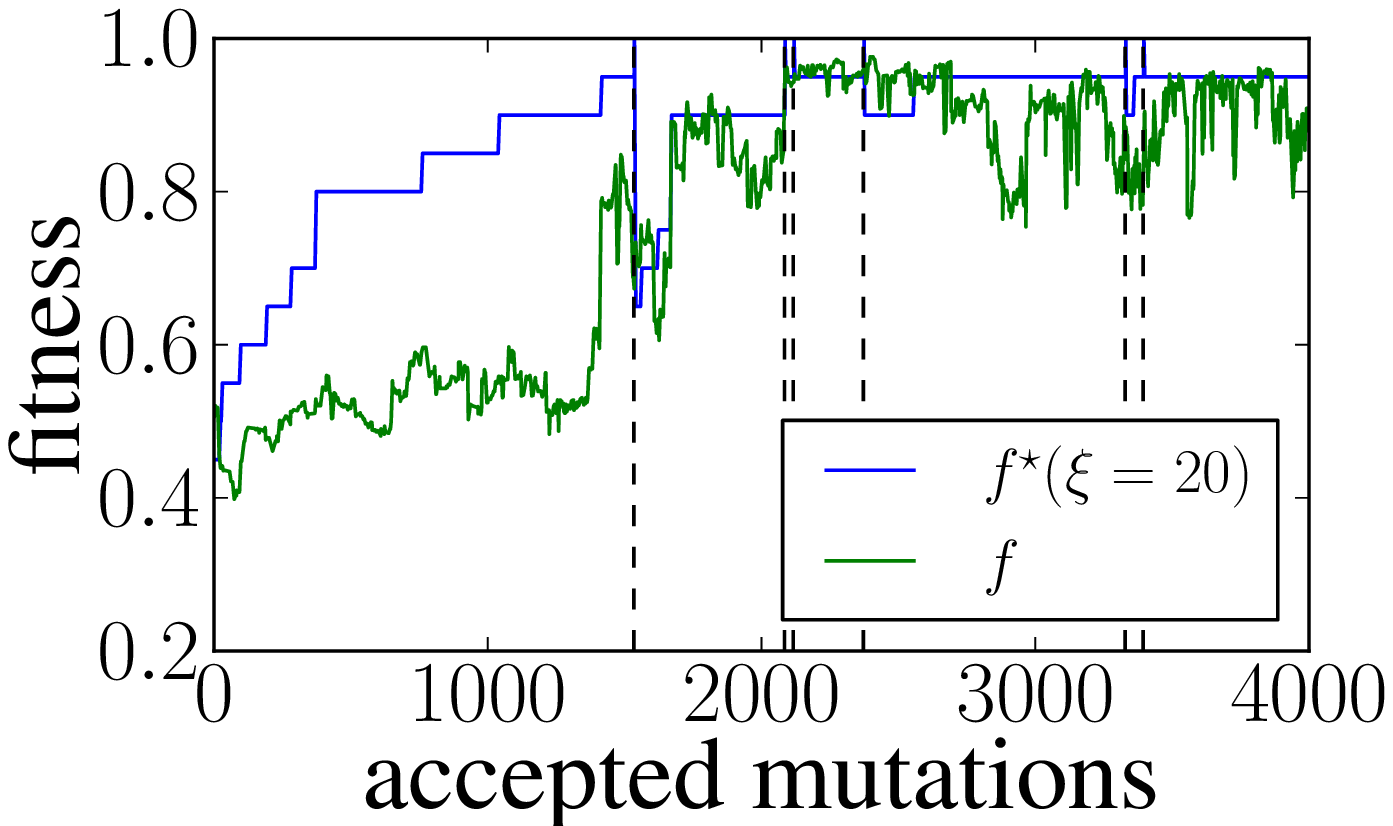}}
  \subfloat[$\xi=40$]{\includegraphics[width=0.49\linewidth]{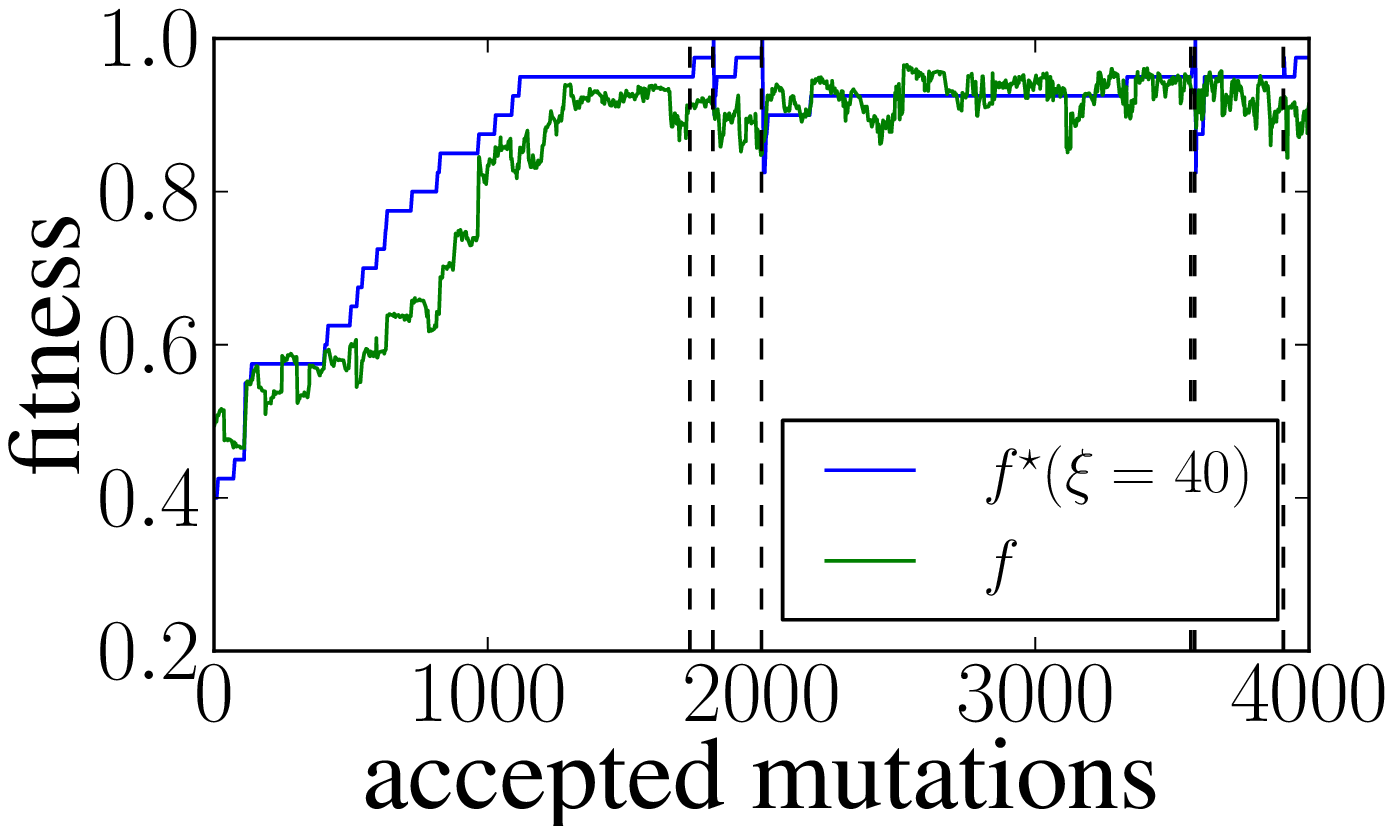}}
  \caption{Evolution via approximate fitness of a network with $N=20$ and $l
    =4.3$, and with sample sizes of $\xi=20$ and $\xi=40$. Vertical lines mark
    the instances when new sets of perturbed nodes were chosen.}
 \label{fig:appr_fit}
\end{figure}

We have investigated the performance of this approximation, as can be seen in
Fig.~\ref{fig:appr_fit}, which compares the approximate and real fitness during
two evolutionary processes applied to the same network, using $f^{\star}$ as the
selection criterion, with sampling sizes of $\xi =20$ and and $\xi=40$. One can
see that the real fitness increases in both cases, and that it fluctuates around
$f^{\star}$, but does not deviate strongly from it. The amplitude of the fluctuation
gets smaller for larger $\xi$.

\subsubsection{Fitness results}

We optimized networks for $N=10,15$ and $l=2,2.5,3,4,6$ as well as for $N=20$
and $l=2,2.5,3$ using the fitness function $f$. Networks with $N>20$ and
$N=20$ with $l=4,6$ were optimized via the approximate fitness function
$f^{\star}$. The number of networks evolved ranged from $10^4$ for $N=10$ to
$800$ networks for larger values of $N$ and $l$.  

The results are shown in Fig.~\ref{fig:fitness}.  A significant fraction of
networks did not reach $f_{\text{max}}$, which can be potentially due to
three reasons:
\begin{enumerate}
\item[(a)] The evolution got stuck in a local fitness maximum.
\item[(b)] The global fitness maximum of the network is smaller than $f_{\text{max}}$.
\item[(c)] The algorithm stopped before the optimization reached $f_{\text{max}}$.
\end{enumerate}
For $N=10$, the fraction of networks that did not reach $f_{\text{max}}$
decreases monotonically with increasing $l$ which indicates that the probability
of reaching $f_{\text{max}}$ increases with the growth of the search space. We
tried to optimize these networks further with a \emph{simulated annealing}
algorithm~\cite{kirkpatrick_optimization_1983} by introducing a probability
$p=e^{-|\Delta f|/T}$ of accepting a negative mutation in order to leave a local
maximum. As this never resulted in better values of fitness, and since all
networks suffered their last positive mutation after approximately $10\%$ of
total running time of the algorithm, we concluded that reason (b) is more
probable than either (a) or (c).  The fraction of networks that did not reach
$f_{\text{max}}$ increases with $l$ for $N=15$ and $N=20$. These networks often
suffered their last positive mutation almost at the end of the optimization run,
and thus one could increase the fitness if we would evolve them further, but it
would take a much longer time for it to increase significantly. However, despite
the fact that many networks did not reach the values of $f_{\text{max}}$, the
deviation from $f_{\text{max}}$ for the final values of fitness are very small,
as can be seen in Fig.~\ref{fig:fitness}. This deviation is worsened if the
approximate fitness is used, as seen in the bottom graph, which can be improved
only if the number of samples $\xi$ is increased, as the change from $\xi=40$ to
$80$ shows.  The total number of mutations needed to evolve the networks can be
as large as a few thousands, for larger $N$ and $l$, and is therefore much
larger than in the work of Szejka and Drossel~\cite{szejka_evolution_2007}. This
is due the fact that the optimization done here is much more restricted, as we
only search through the space of possible update functions whereas in
\cite{szejka_evolution_2007} both the topology and dynamics were allowed to
change, and there was no particular trajectory imposed on the system.

\subsection{Update functions}

We evaluated the frequency of the possible update functions that occur
in the optimized networks.  Let us first discuss the functions with
$k=2$. Before and after optimization for robustness, the distribution
is almost entirely dominated by the eight canalizing functions that have
three bits of one type and one bit of the other type in the truth
table. The reasons for the dominance of these functions were explained
in~\cite{peixoto_boolean_2009}. For functions with larger $k$, we
evaluated the homogeneity $d$, which is equal to the number of entries in
the truth table that have the minority bit. 
\begin{figure}[t]
\centering
 \subfloat[$k=3$, before evolution]{\includegraphics[width=0.33\linewidth]{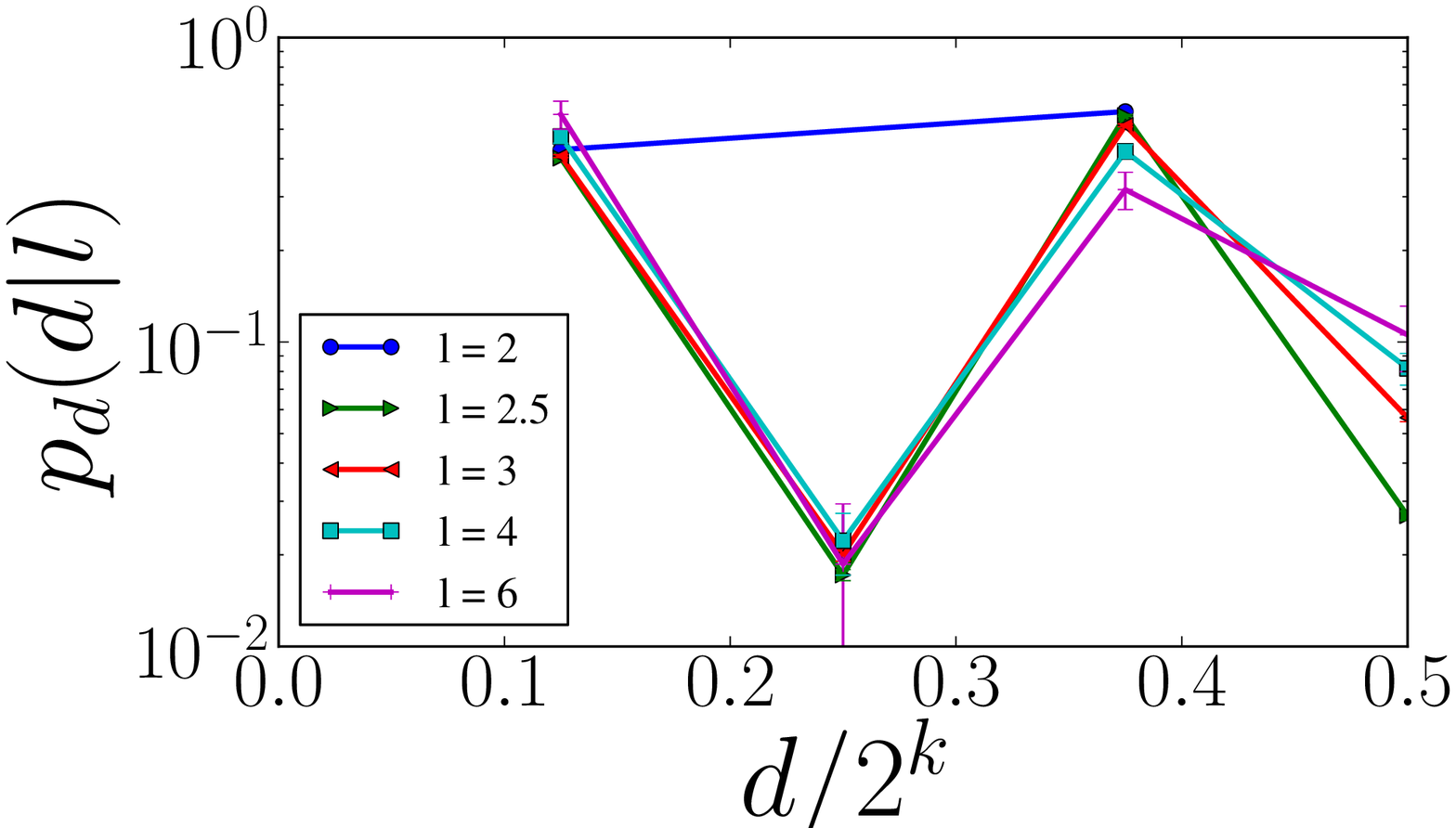}}
 \subfloat[after evolution]{\includegraphics[width=0.33\linewidth]{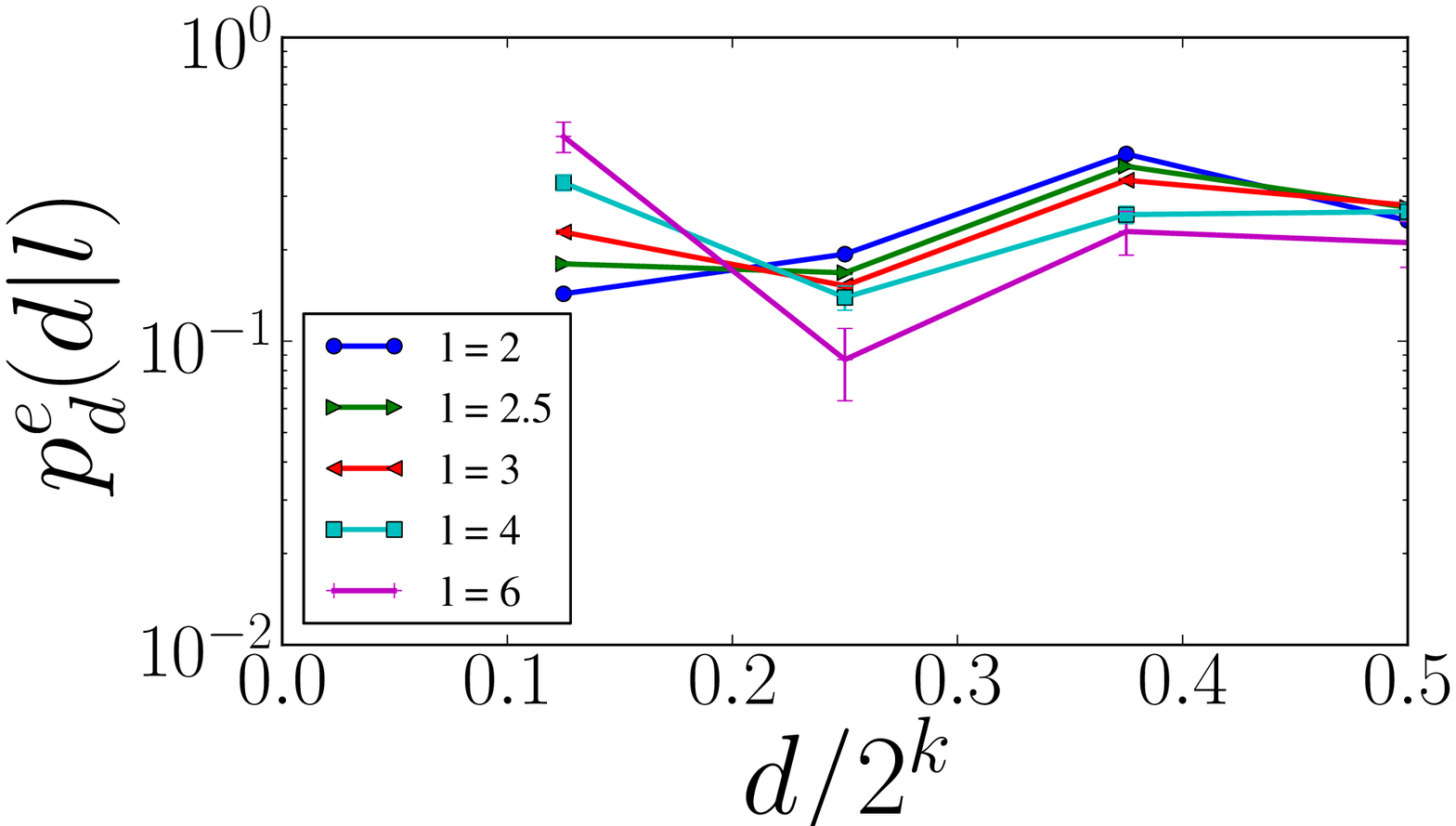}}\subfloat[after homogenization]{\includegraphics[width=0.33\linewidth]{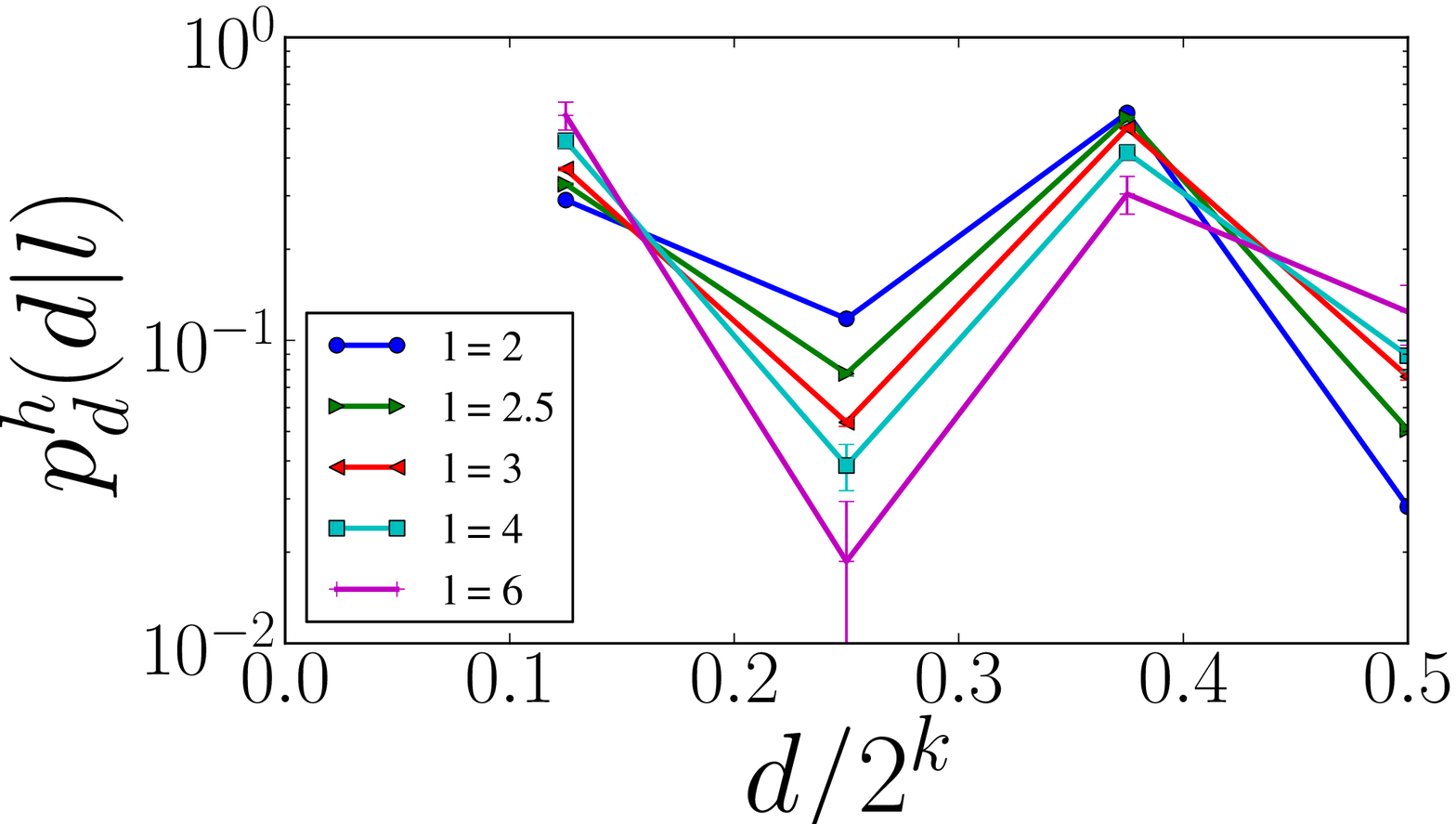}}\\
 \subfloat[$k=4$, before evolution]{\includegraphics[width=0.33\linewidth]{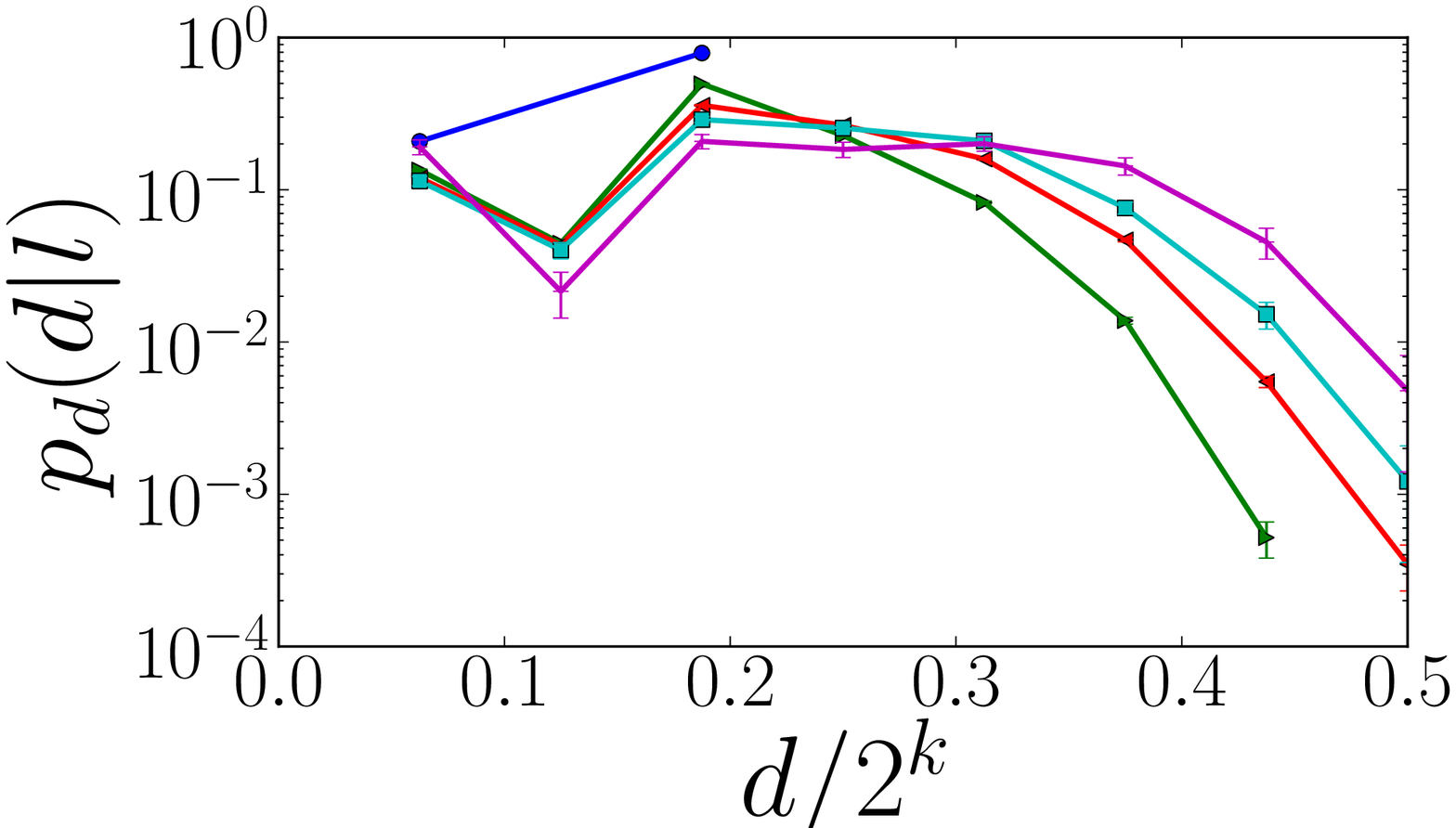}}
 \subfloat[after evolution]{\includegraphics[width=0.33\linewidth]{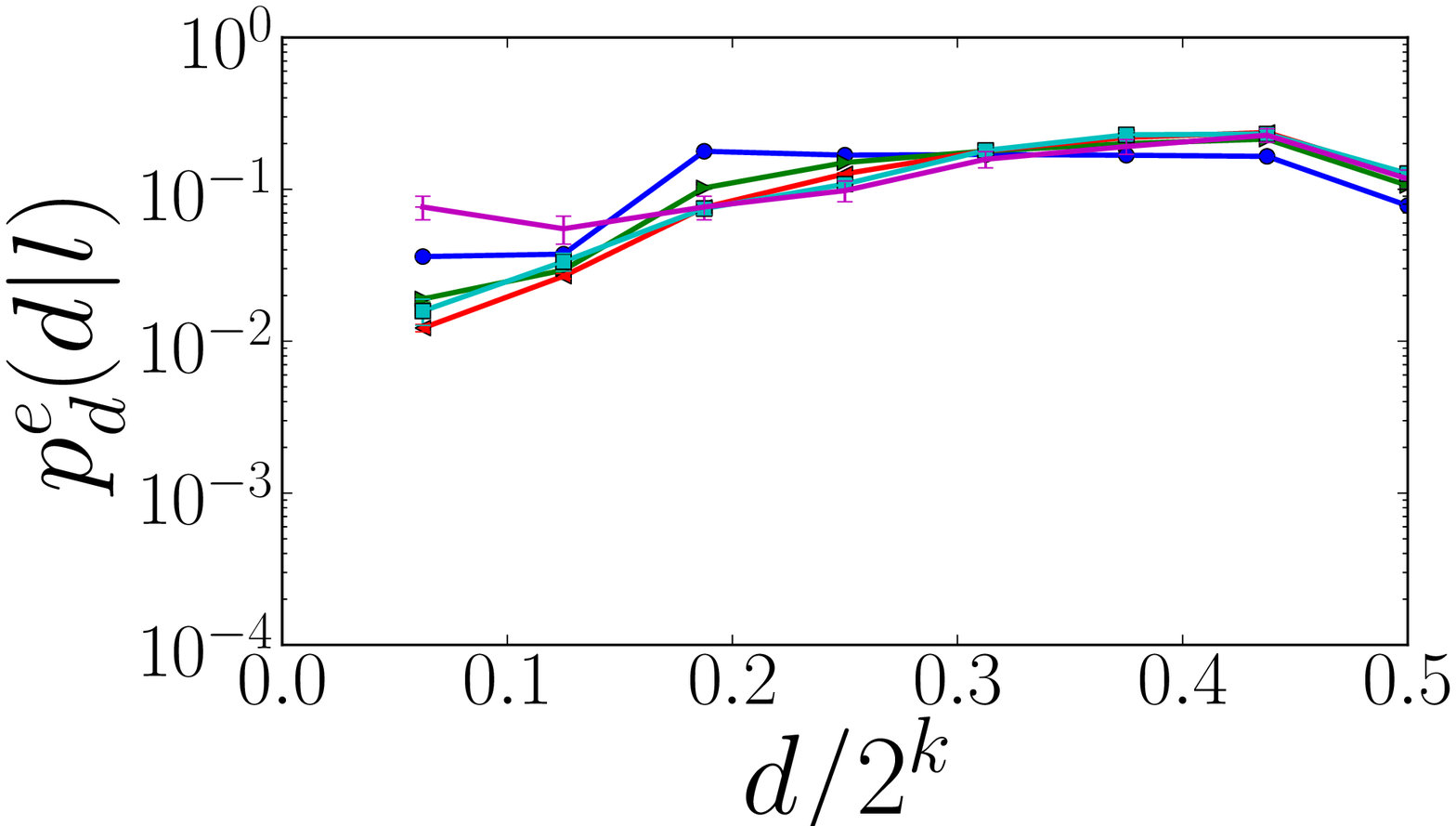}}\subfloat[after homogenization]{\includegraphics[width=0.33\linewidth]{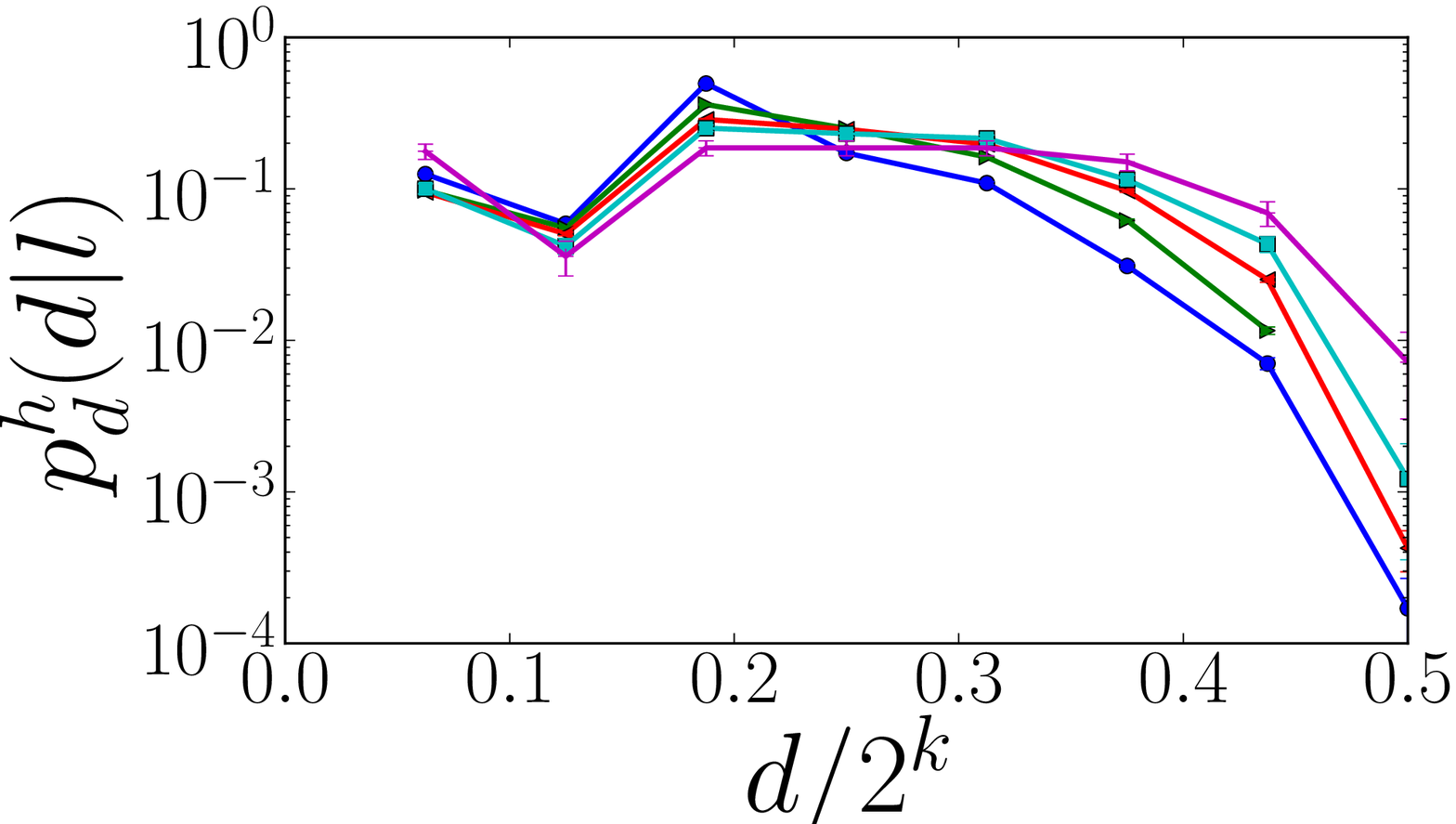}}\\
 \subfloat[$k=5$, before evolution]{\includegraphics[width=0.33\linewidth]{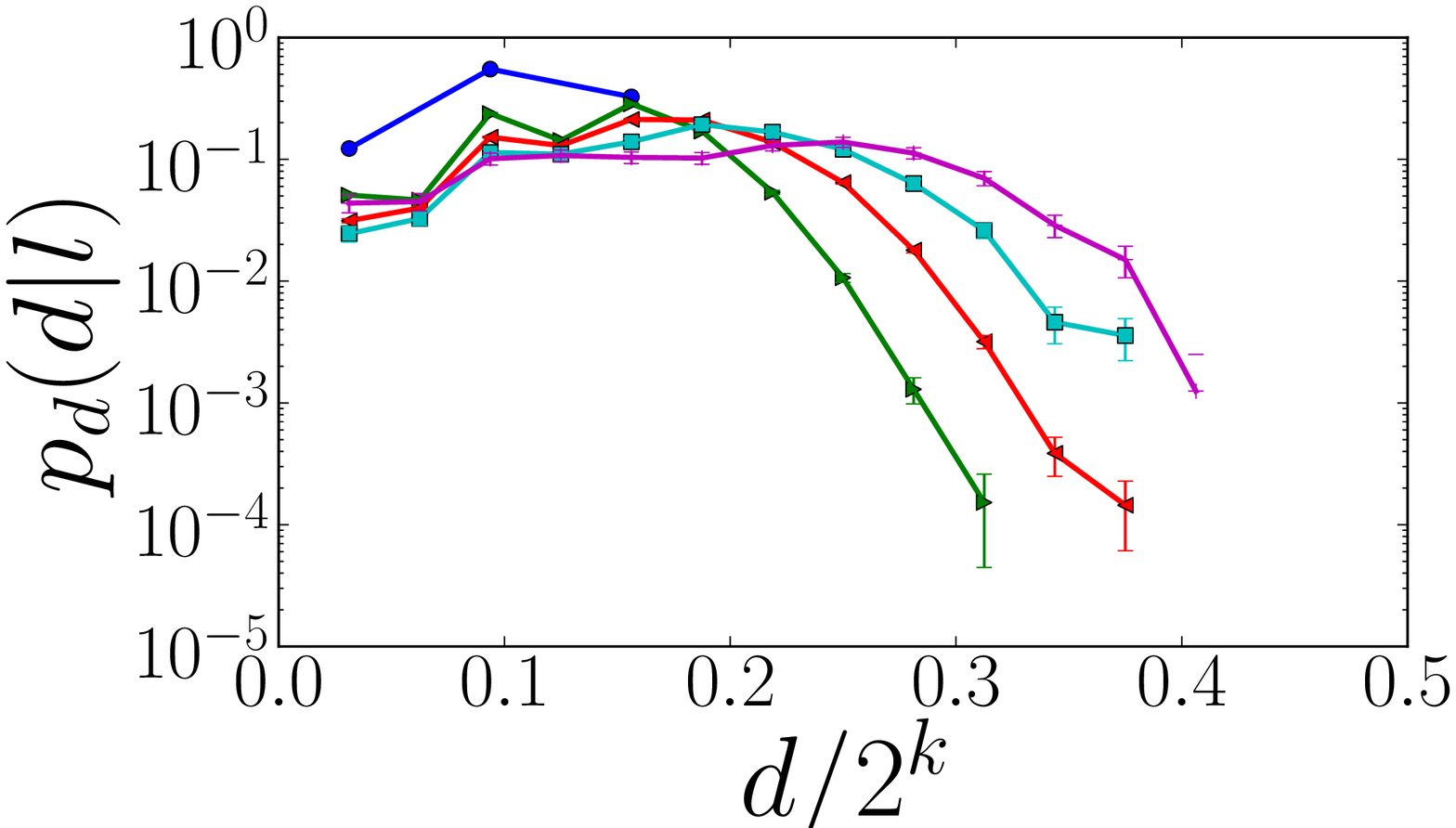}}
 \subfloat[after evolution]{\includegraphics[width=0.33\linewidth]{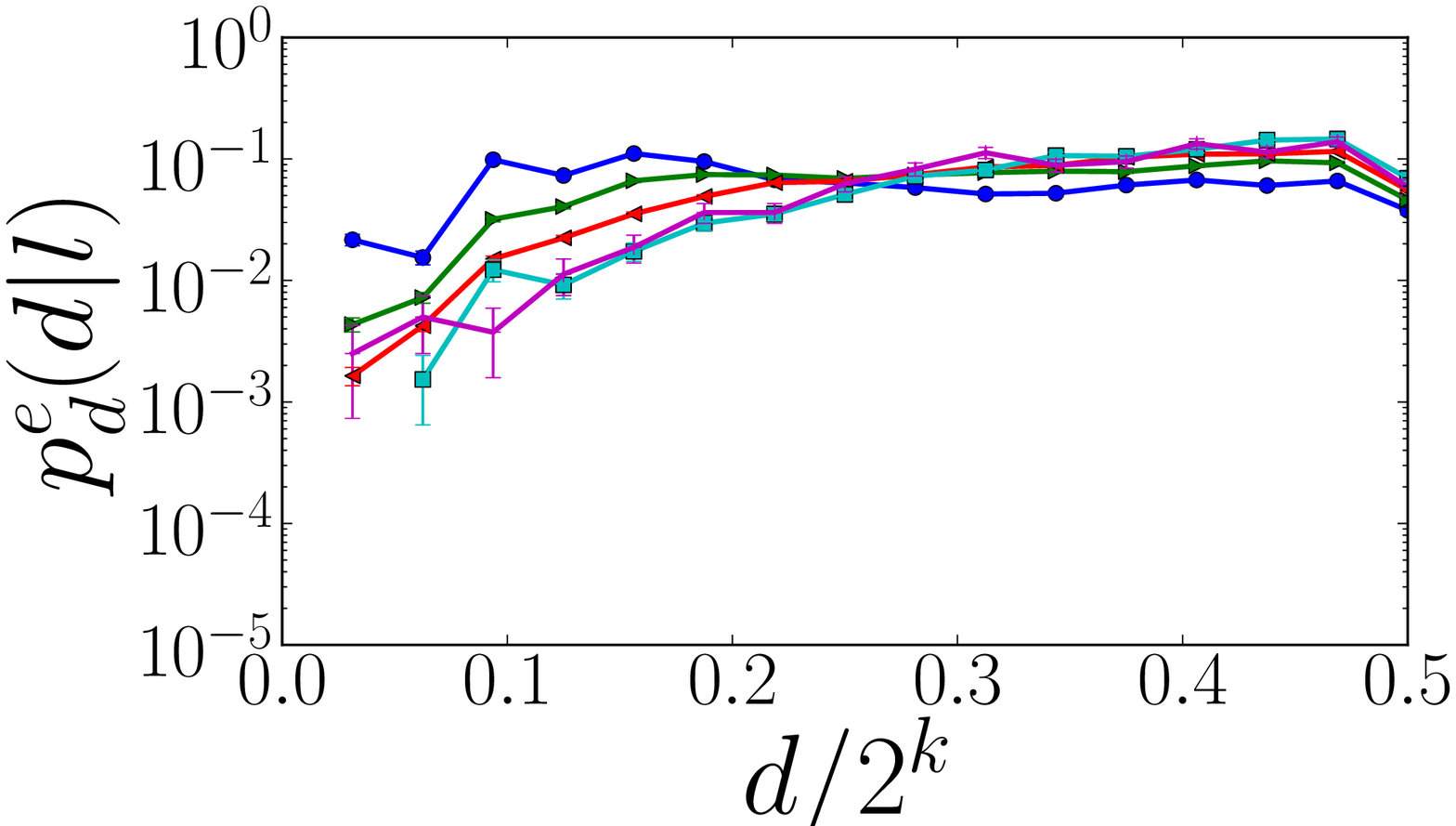}}\subfloat[after homogenization]{\includegraphics[width=0.33\linewidth]{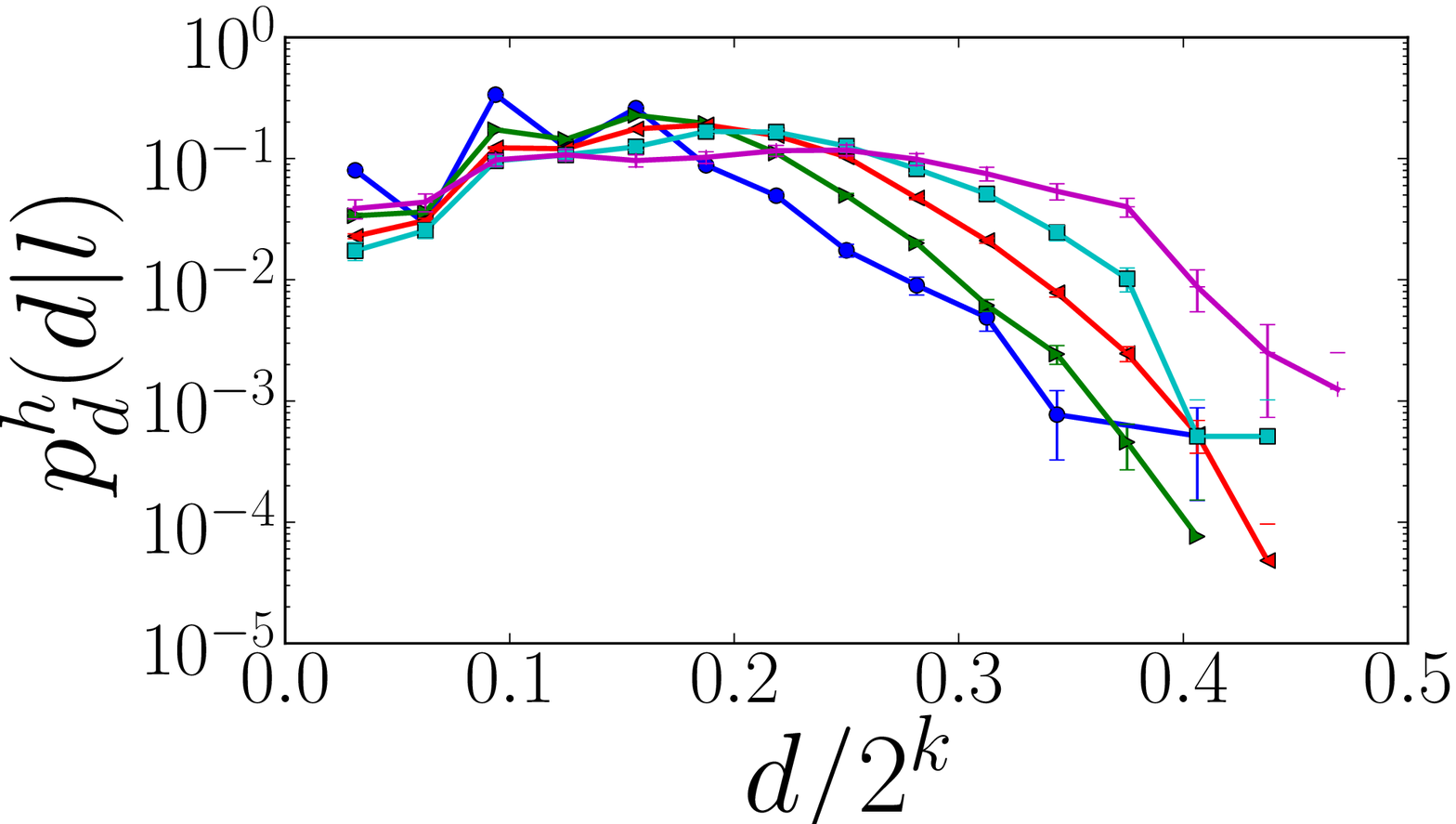}}
 \caption{Distribution of functions with different values of $d$, for different
   $l$ and $N=10$ before the evolutionary optimization (left), after evolution
   (middle), and after homogenization (right).}
 \label{fig:hom_dis}
\end{figure}
Fig.~\ref{fig:hom_dis} shows the distribution of $d$ before and after the
optimization process. Before the optimization process, functions with smaller
values of $d$ dominate. After the process, the number of functions with higher
values of $d$ is significantly larger. This means that the distribution of
functions has become more random, as there exist many more functions with larger
$d$, their number being $2 {2^k \choose d}$ for $d<2^{k-1}$, and ${2^k \choose
  d}$ for $d=2^{k-1}$.

In order to investigate whether the differences in homogeneity are a fundamental
property of the optimized networks or an artifact of neutral mutations, we tried
to decrease the values of $d$ while retaining the value of fitness. To achieve
this, we let the evolutionary algorithm continue from the final configuration,
with the modification that a mutation is only accepted if it simultaneously does
not lower the fitness and increases the homogeneity of the randomly chosen
update function. This was done for the evolved networks with $N=10$ and
$N=15$. The distribution of $d$ after homogenization is shown in the right
column of Fig.~\ref{fig:hom_dis}.  It can be clearly seen that the shift to less
homogeneous functions can be reversed to a large degree, except for $l=2$. This
means that the increase of the values of $d$ during the evolutionary process is
mainly due to neutral mutations.  The fact that it is possible to homogenize the
update functions after reaching the global optimum $f_{\text{max}}$ gives an
insight into the fitness landscape: The global optimum is a plateau, instead of
isolated peaks, on which the networks can move via neutral mutations, similar to
what was found in~\cite{szejka_evolution_2007}.

\subsection{State space}
\begin{figure}[htbp]
\centering
 \subfloat[before evolution]{\includegraphics[width=0.32\linewidth]{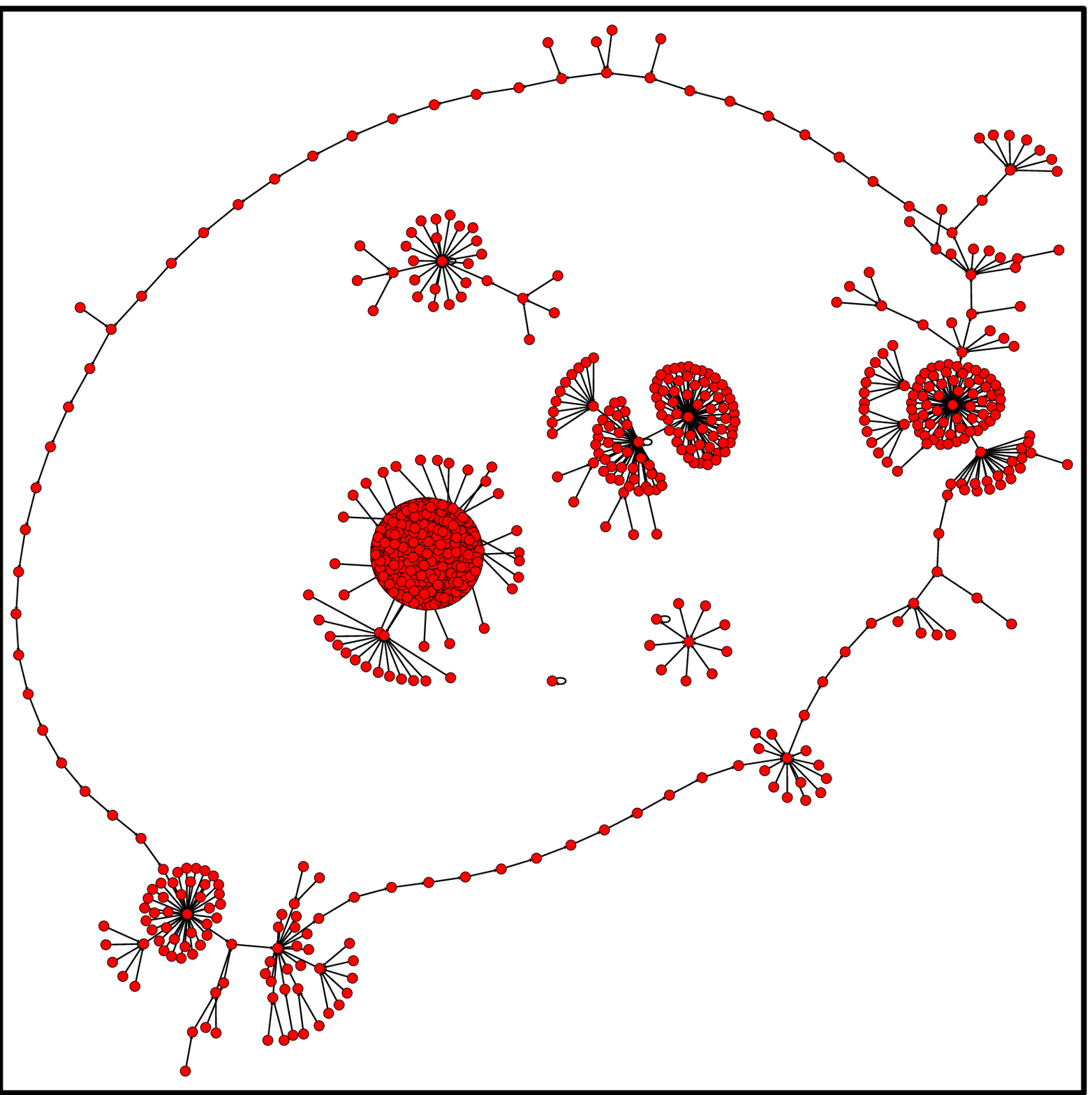}}
 \subfloat[after evolution]{\includegraphics[width=0.32\linewidth]{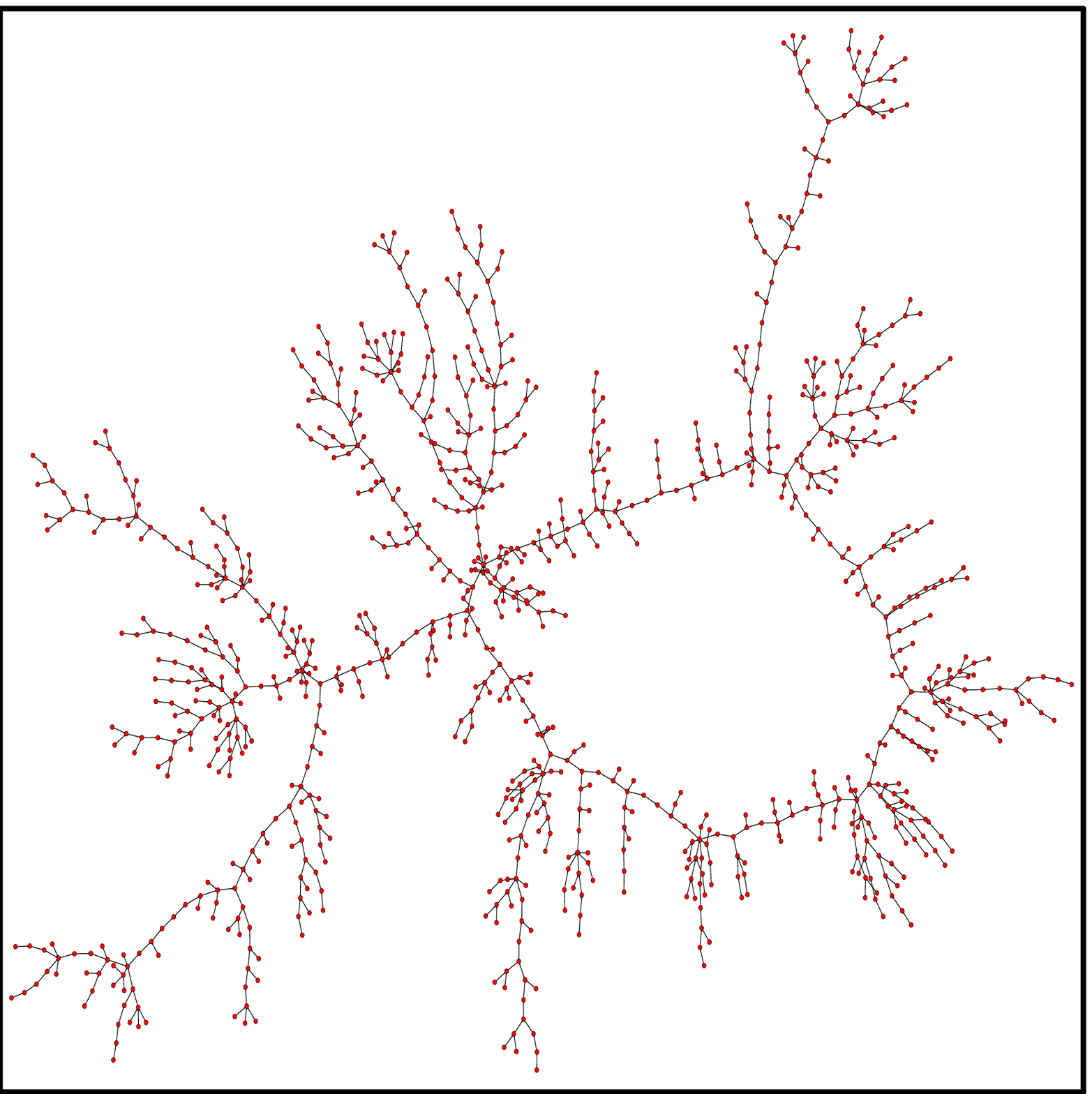}}
 \subfloat[after homogenization]{\includegraphics[width=0.32\linewidth]{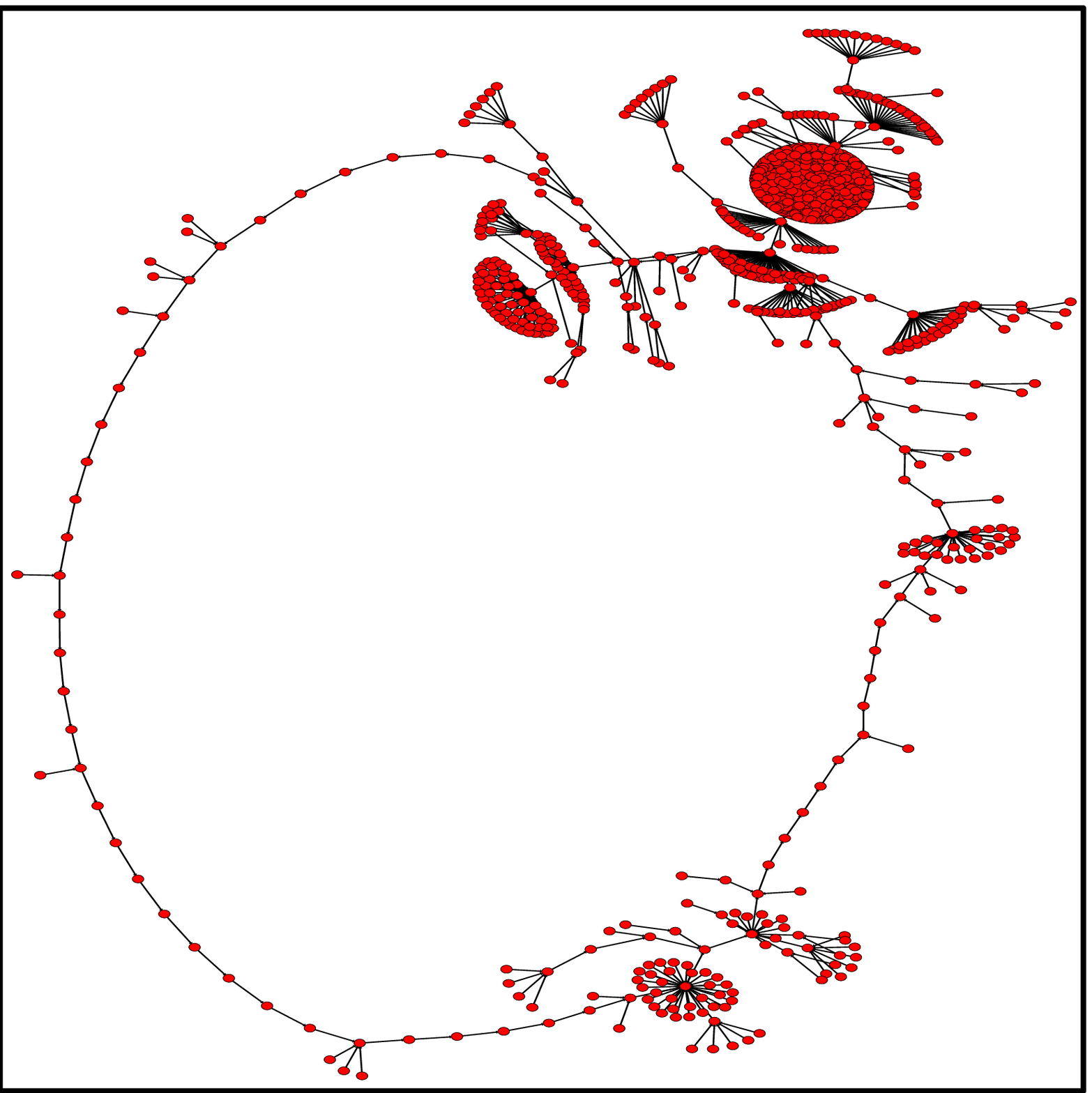}}
 \caption{Influence of the evolution and homogenization on the state space of a network with $N=10$ and $L=68$.}
 \label{fig:sp_evolution}
\end{figure}

Lastly, we investigated the influence of the optimization and
homogenization processes on the state space of the minimal
networks. We evaluated the entire state space for optimized networks
of size $N=10, 15$,  and we sampled the state space for $N=20$, under
parallel update. In particular, we enlisted the attractors and the
sizes of their basins of attraction (i.e., the number of states
leading to the attractor).
\begin{figure}[htbp]
\centering
\includegraphics[width=0.79\linewidth]{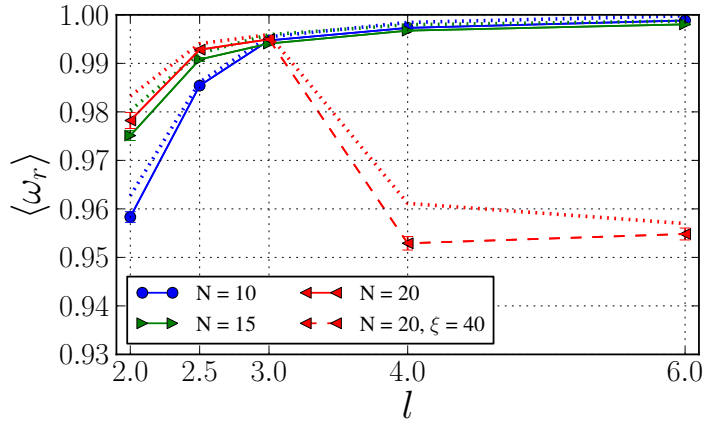}
\caption{Average basin size $\left<\omega_r\right>$ of the reliable attractors
  for different values of $N$ and $l$. The dotted line shows the averages after
  the evolutionary process. The dashed lines represents networks evolved via the
  approximate fitness.}
\label{fig:state_space}
\end{figure}
As expected, the optimization process increases the average basin size of the
reliable trajectory $\omega_r$, as can be seen in Fig.~\ref{fig:state_space}.
Fig.~\ref{fig:sp_evolution} shows the state space of a typical network with
$N=10$ nodes and $l = 6.8$, before and after evolution, and after homogenization.
Before evolution, the state space is divided into six basins of
attraction, five of which belong to fixed points, and one being the
basin of the reliable attractor. The network has an unevolved fitness
of $f\approx 0.64$. The short transients of $T\approx1.3$ steps on
average indicate that the system resembles an RBN in the frozen phase.
After the evolutionary process, the basin of the reliable trajectory
occupies the entire state space, leading a fitness of $f=1$. The
dynamics are less frozen, with the average transient time to the
attractor having increased to $T\approx 10.1$. After homogenization,
the transient time has decreased to $T\approx 2.9$.

\section{Conclusion}\label{sec:conclusion}
We have shown that there exists a large ensemble of minimal Boolean
networks that show reliable and robust dynamics. The networks are
minimal in the respect that the number of connections of a node is not
larger than necessary for obtaining a desired reliable trajectory. A
reliable trajectory is an attractor of the dynamics of the network
that does not change when the update schedule is changed or
randomized.  This means that under parallel update, at each time step
only one node changes its state. The reliable trajectories were chosen
at random, given a fixed average number of flips per node. High
robustness was achieved by using an evolutionary algorithm that
modifies the update functions and that accepts only those changes that
do not decrease robustness.  For all investigated parameter sets, we
obtained networks with a robustness close to 100 percent. The set of
update functions associated with the final robustness value is not
unique, but can be varied over a broad range of homogeneity
values. (Homogeneity is quantified by the average number of bits in
the truth table of the update function that differ from the majority
bit.) The state space of the resulting networks is dominated by the
basin of attraction of the reliable trajectory. 

Dynamical reliability and robustness to noise are important features
of biological networks, such as gene regulation networks. While the
networks constructed by our procedure are random in many respects and
still far from the very specific networks found in biological systems,
our study shows that there exist many solutions to the task of
constructing such networks. 

\section*{References}

\bibliographystyle{unsrt}
\bibliography{bool_references.bib}
\end{document}